# Optical secret sharing with cascaded metasurface holography

Philip Georgi*,[1], Qunshuo Wei*,[2], Basudeb Sain[1], Christian Schlickriede[1], Yongtian Wang[2,†], Lingling Huang[2,‡] and Thomas Zentgraf[1, §]

[1] Department of Physics, Paderborn University, Warburger Straße 100, 33098 Paderborn, Germany
[2] Beijing Engineering Research Center of Mixed Reality and Advanced Display, School of Optics and Photonics, Beijing Institute of Technology, 100081, Beijing, China

Secret sharing is a well-established cryptographic primitive for storing highly sensitive information like encryption keys for encoded data. It describes the problem of splitting a secret into different shares, without revealing any information about the secret to its shareholders. Here, we demonstrate an all-optical solution for secret sharing based on metasurface holography. In our concept, metasurface holograms are used as spatially separable shares that carry an encrypted message in form of a holographic image. Two of these shares can be recombined by bringing them close together. Light passing through this stack of metasurfaces accumulates the phase shift of both holograms and can optically reconstruct the secret with high fidelity. On the other hand, the holograms generated by the single metasurfaces can be used for identifying each shareholder. Furthermore, we demonstrate that the inherent translational alignment sensitivity between the two stacked metasurface holograms can be used for spatial multiplexing, which can be further extended to realize optical rulers.

Introduction

Over the years, several approaches have been developed to protect data from theft and tampering. A common way to enhance data security is information encryption. Nowadays, cryptographic methods are widely used for security-related problems concerning the protection of intellectual property rights and product authentication. Invented by Adi Shamir and George Blakley in 1979, secret sharing solves the cryptographic task of splitting a secret among multiple shareholders so that it can only be reconstructed when a sufficient number of shares are combined together(1, 2). In doing so, the corruption of a single shareholder does not leak any information about the shared secret. In another type of security application, optical holography has appeared as a crucial security element against counterfeiting with several important applications in identity cards, passports, and banknotes(3-5). Among the different available holographic techniques, meta-holography thrives with its inherent flexibility and compactness, which allows for dense storing and encrypting of optical information into spatially distributed phase and amplitude patterns(6, 7).

The design flexibility of metasurfaces stems from the huge variety of possible nanostructure shapes and can be used to access multiple degrees of freedom (DOF) per pixel. For metasurface holography, these additional DOF offer the possibility of realizing novel multiplexing schemes for improved information security, where different holographic images are encoded along multiple optical dimensions(*8-16*). However, so far, metasurface multiplexing frameworks were usually one-layer solutions, which only allow for the access of different information channels, but not for the physical splitting of these information channels across multiple shareholders. In this context, multilayer layouts offer an alternative as they allow for the physical separation of layers and their inherent information.

[†] Email: wyt@bit.edu.cn
[‡] Email: huanglingling@bit.edu.cn
[§] Email: thomas.zentgraf@uni-paderborn.de

Previously, multilayer metasurface layouts were often used to achieve new functionalities or to increase the performance of existing optical metasurface elements. This includes the demonstration of high wavelength selectivity(*17, 18*), multispectral(*19, 20*) and varifocal(*21*) metalenses, metasurface retroreflectors(*22*), holography with asymmetric transmission(*23*), circular polarization filters(*24*), multicolor holography(*25, 26*), image differentiation(*27*), and quantitative phase microscopy(*28*) as well as pattern recognition in an all-optical machine learning framework(*29*). The advantage of multilayer phase mask systems or diffractive deep neural networks is that they are highly customizable as they gradually change the state of light within each layer(*29, 30*). Some demonstrations based on multiple phase profiles have been proposed for watermarking and optical encryption(*31-35*). However, independent holographic images of each single phase-profile were not obtained. With advances in design and algorithmic techniques these multilayer systems can be developed towards innovative holographic security features(*36*).

## Results

Here, we experimentally demonstrate a cascaded metasurface framework that can be used to split and share encrypted holographic information across multiple metasurface layers (Figure 1). In our optical secret sharing scheme, we use a set of nanostructured optical metasurfaces as 'shares'. Each of them contains an encoded phase-only Fourier hologram which can be reconstructed in the far-field upon illumination with circularly polarized light. These single-layer images serve as a unique identifier for each metasurface. Meanwhile, when the two metasurfaces are stacked 100 µm apart, the illumination of the cascaded configuration creates a new holographic image that is distinct from their two single-layer holograms. This cascaded holographic image can be used as an optical secret that is only revealed if both metasurfaces are stacked. The concept can be extended to a larger set of shares where only the combination of all of them will reveal the stored secret. To achieve this behavior, we use an iterative gradient optimization approach that utilizes the 'automatic differentiation' feature of modern machine learning libraries(*37*).

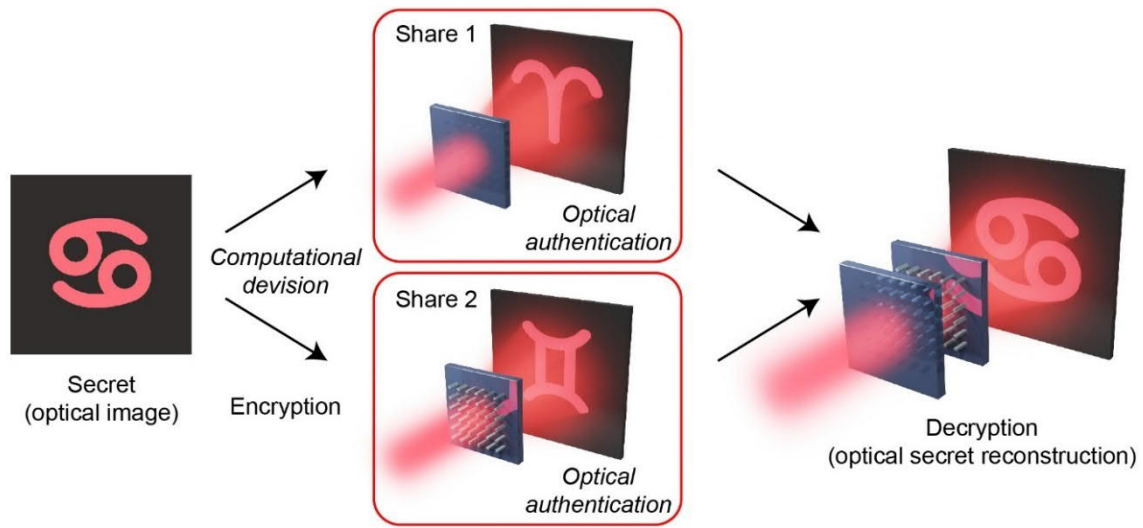

Figure 1: Conceptual illustration of the holographic secret sharing via cascaded metasurfaces. By design, both metasurfaces (shares) contain a Fourier hologram, which reconstructs an image in the visible when they are illuminated with circularly polarized light. At the same time, the illumination in the cascaded configuration creates a completely different image. While the secret of the cascaded image is shared across two metasurfaces, the single metasurface holographic images can be used as unique identifiers for the shareholders.

In a simple picture, the cascaded hologram's functionality can be understood as follows: Light passing through the cascaded arrangement accumulates the phase delays of both layers. In an idealized case, where both metasurfaces are virtually placed on top of each other, this would correspond to adding up both phase masks pixelwise. Considering that the same cascaded phase mask can be built up through combinations of different single phase-masks, it is possible to share the 'secret' of the cascaded image across two metasurface layers without revealing any information about the shared secret within the single-layer images. Meanwhile, by adding up both phase profiles pixelwise, the phase relations of the single phase-masks are no longer preserved which allows for the multiplexing of single and cascaded images.

For our demonstration, we use the concept of the Pancharatnam-Berry-phase (PBP) to realize the metasurface phase masks(*38, 39*). Thus, the metasurfaces are made of silicon nanofin structures which are designed as local half-wave plates. According to their individual orientation, these nanostructures introduce a spatial variant phase delay into the cross circular polarization state of the transmitted light. The PBP has not only the advantage of being quite robust with regard to fabrication and wavelength, but it also allows for filtering unwanted light through the usage of polarization optics. We fabricate two distinct metasurface holograms on one glass substrate (Metasurface Set A) and three additional metasurface holograms on a second glass substrate (Metasurface Set B). This allows for six different combinations in a cascaded configuration. Details about the design and fabrication can be found in the Methods.

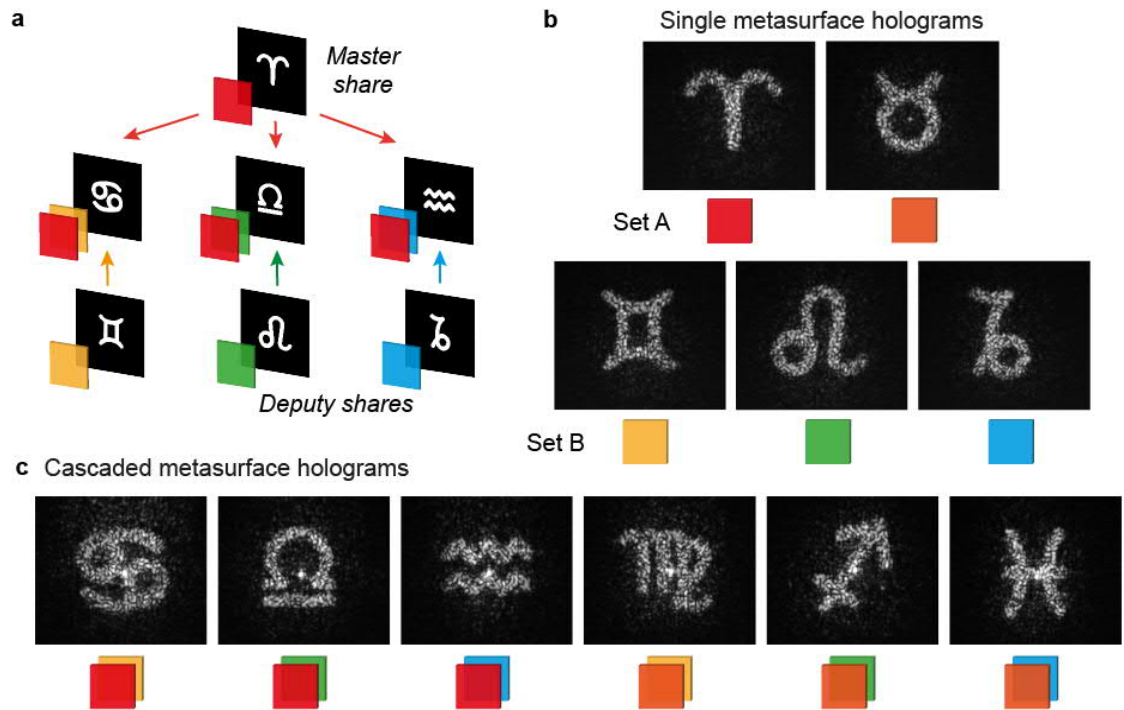

Figure 2: Holographic encryption scheme. a Illustration of the concept using a "Master Share" metasurface hologram in combination with three distinct "Deputy Shares". For the demonstration, we fabricate two distinct Master Share metasurface holograms on one substrate (Set A) and three Deputy Share metasurface holograms on a second substrate (Set B). Within every single-metasurface of both sets and each stacked combination, a distinct Fourier hologram of an astrological symbol is encoded. b Measured single-metasurface holographic images for a wavelength of 740 nm. For better visualization, every metasurface hologram is represented by a unique color. c Measured cascaded-metasurface holographic images for the various combinations of the metasurfaces of both sets for the same wavelength. The cascaded images nearly achieve the image quality of their single-layer counterparts.

The experimentally reconstructed images of the encoded astrological symbols for the single metasurface holograms and their combinations are shown in Figure 2. We observe that all five metasurface holograms reconstruct their individual single-layer image in high quality with low background noise (Fig. 2b). Meanwhile, cascading the metasurface holograms from both sets results in six reconstructed images with a similar quality compared to their single-layer counterparts (Fig. 2c). However, the cascaded image quality strongly depends on the alignment and deteriorates for small translational in-plane shifts between both metasurfaces. This behavior can be understood by considering that the appearance of the cascaded images requires correct pixel matching between both metasurfaces. Thus, we observe a breakdown of the cascaded hologram image when the translational mismatch exceeds the pixel size of 5 μm (see Supplementary Information).

Furthermore, all images are clearly visible without any apparent cross-talk (single metasurface images are not visible in the cascaded case). The low cross-talk can be explained by the fact that the second metasurface completely wipes out any regularities of the first metasurface's phase profile and vice versa. In the limit of zero distance between the metasurfaces, this can be understood as follows: Since the Fourier transform of a product is the convolution of its Fourier transformed elements, the reconstructed cascaded image is effectively the convolution of the two single images' complex electric fields. In the vast majority of cases, this convolution operation does not preserve any visible information from the single images to the cascaded image. Therefore, we do not observe any image artifacts of the single-layer holograms within the cascaded cases. The low cross-talk is in particular important for optical secret sharing applications where the single metasurface hologram should not reveal any information of the secret.

In our example, we chose to use two holograms in Set A and three holograms in Set B. However, the concept is general and the number of encoded images can be increased further by adding more metasurfaces to both sets or by increasing the number of sets. Note that the two sets of meta-holograms (Master Set A and Deputy Set B) can be combined arbitrarily. All individual single-layer metasurface holograms and the various combinations of them carry their holographic information. With the entire series of combinations, the complete secret set can be revealed. Based on such scaling of information, we need to recognize that the achievable image quality is fundamentally limited by the ratio of encoded images to available phase masks. A good scaling strategy is to have a master set with a small number of shares and a deputy set with a larger number of shares. In such a scenario the image quality primarily depends on the number of shares in the master set. Note, that a decrease in image quality due to a high number of shares can be compensated by increasing the number of pixels in each hologram.

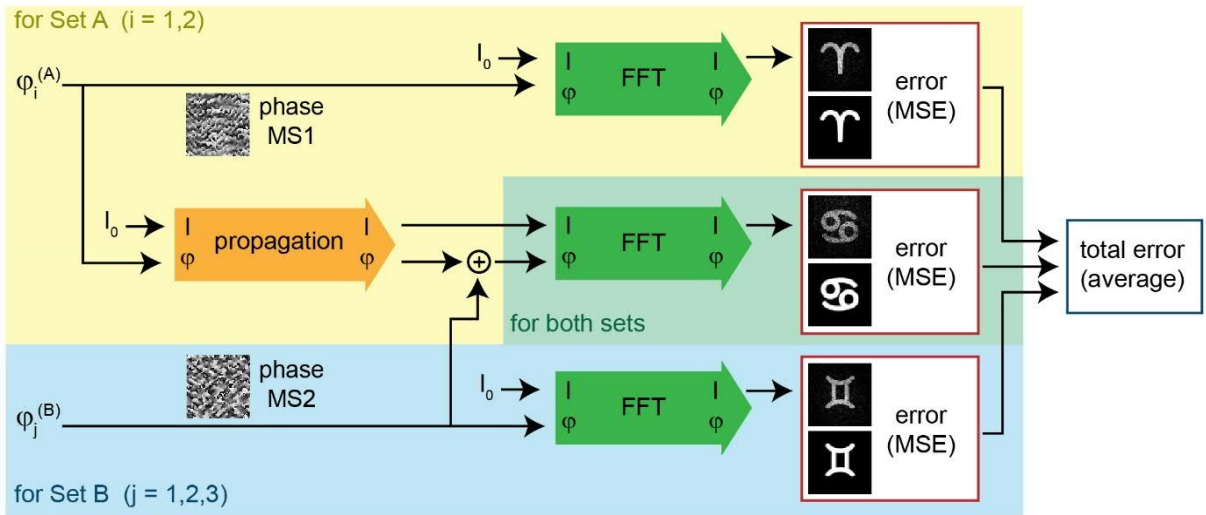

Figure 3: Flow chart of the forward pass for the gradient optimization scheme. The metasurface-phase masks (MS1 & MS2) used in the cascaded configuration were designed using a gradient optimization scheme. The flow chart shows the physical forward pass within the gradient optimization loop, which assigns an error or fitness value to a given set of phase masks. The algorithm performs a fast Fourier Transformation (FFT) for all single and cascaded images (Set A and B) and then compares the calculated images with target images by using the mean squared error (MSE). For the light propagation between the two metasurface layers in the cascaded case, we use the angular spectrum method with zero padding. In the optimization loop, we try to find a set of phase masks that minimizes the total error $\alpha$. This is accomplished by calculating its gradient $\nabla_\varphi \alpha(\varphi)$ with automatic differentiation and using a gradient optimizer.

To obtain high-quality images, the design algorithm for the holograms plays an important role. Here, we treat the metasurface design process as an optimization problem in which we try to minimize the differences between the target images and the calculated images resulting from the simulated light interference. Figure 3 schematically illustrates the calculation of the total error function, which benchmarks the image qualities for a given set of metasurface phase masks. The error calculation is a two-step process, which involves both the physical recreation of the image formation for a given phase mask set as well as an image comparison to the target images. It is important to note that the entire forward calculation has to be explicitly performed within an *automatic differentiation framework* in order to be traced for the gradient calculation. The gradient $\nabla_\Phi \alpha(\Phi)$, which is the derivative of the total error $\alpha$ with regard to all phase mask values $\Phi = \prod_{i=1,2} \varphi_i^{(A)} \times \prod_{j=1,2,3} \varphi_j^{(B)}$, is then used within a gradient optimization scheme, which converges towards a local optimum (for more details see Supplementary Information). The actual secret splitting is achieved implicitly within the optimization framework. By providing a target for all single-layer configurations we create an algorithmic incentive to hide the cascaded image for a single shareholder.

Our developed algorithmic approach is an efficient and straightforward design method for the demonstrated holographic secret sharing framework. It should be noted that the gradient optimization with regular numerical gradient calculation becomes infeasible due to the high number of degrees of freedom. Moreover, alternative approaches based on the widely used Gerchberg-Saxton algorithm might not inherently guarantee convergence and can struggle to encode a larger number of holograms(*40, 41*).

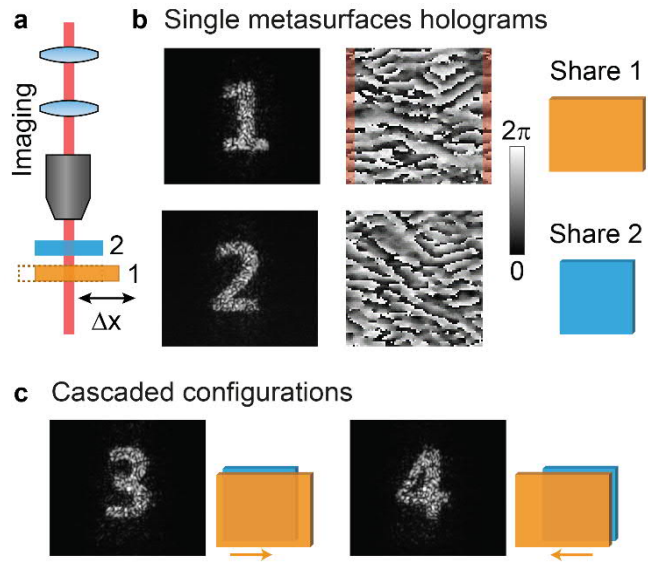

Figure 4: Translational multiplexing with cascaded holograms. a Schematic view of the experiment containing both metasurface Shares 1 and 2. The high alignment sensitivity for the cascaded image reconstruction allows for the implementation of translational multiplexing by shifting metasurface 1. b The two metasurfaces are designed and fabricated to contain one single Fourier hologram each as well as two cascaded holograms. The phase pattern of the two holograms are shown in the middle. Note that the metasurface for Share 1 has been periodically expanded to ensure the same spatial overlap after the translation. c A spatial translation of only five pixels (25 μm) in the cascaded case results in a different holographic image reconstruction. The images with the numbers 1-4 show the experimentally measured Fourier images.

To illustrate the potential of cascaded metasurface holography for optical secret sharing, we additionally introduce the concept of translational multiplexing for cascaded holograms. In the previous example, we showed that the shared secret cannot be revealed if the position of the two metasurfaces is mismatched, resulting in the precise alignment requirement. We take advantage of this feature by utilizing the translational movement as a new multiplexing dimension (additional DOF). For that, we encode multiple images across different relative translational positions. Thus, different holographic images can be reconstructed if the metasurface holograms are slightly repositioned against each other.

As proof of principle, we demonstrate here a case where two different translational positions result in two different images for the cascaded hologram (Figure 4). The single metasurface holograms (Shares 1 and 2) contain the encoded images of the numbers '1' and '2'. At the same time, the images of the numbers '3' and '4' appear in the cascaded arrangement for two distinct overlap positions which lie five pixels (25 µm) apart along the x-axis. As the translational movement naturally changes the overlap area between both metasurfaces, we periodically expand the holographic phase mask of Share 1 into a rectangular shape to ensure a square-shaped overlap in both cascaded configurations. Thus, the two cascaded images appear for the alignment of Share 2 along with one of the two edges (left and right) of Share 1. Since we previously observed that the cascaded holographic image quickly deteriorates for a translational movement of one pixel, we expect a negligible cross-talk here for the implemented translational shift of five pixels. This expectation is indeed confirmed by the measurement results in Figure 4c (for additional information see Supplementary Information). In terms of practical applications, the translational multiplexing concept has the potential of being used as an optical ruler. There, cascaded images can be encoded into a discrete number of equidistant real space positions that digitally display the current overlap position as a holographic image. Here, the metasurface holograms have the advantage that micrometer pixel sizes and small spacings can be achieved what in turn would allow for an optical ruler with a very high spatial resolution even below the diffraction limit of light.

Discussion

Our demonstrated stacking of multiple holographic metasurfaces and using them in a cascaded way allows for a unique optical secret sharing and encryption platform. The concept is scalable, both in terms of the stack size and the number of keys. It can be also altered into a ($t$, $n$)-threshold scheme, where already a subset of shareholders ($t$ out of $n$) can reconstruct the secret. Additional complexity and security can be added by using complex amplitude and phase holograms. Because of their compact shape and their design flexibility, metasurfaces are naturally suited for the realization of cascaded configurations. The high sensitivity for the spatial alignment is also perfectly suited for holographic seals, where two metasurfaces are slightly glued together and the restoration of the cascaded hologram would become extremely challenging after both metasurfaces have been separated. For realizing a reprogrammable encryption scheme, one can combine a metasurface as the main share, with a spatial light modulator as the deputy share. By generating different deputy shares with the spatial light modulator and using it for the illumination of the metasurface, different secrets can be revealed.

Methods

*Experiment and sample fabrication*

All holographic images were captured by projecting the Fourier Space of the metasurface plane on a CMOS camera (EO USB 2.0 CMOS) with the usage of a microscope objective (Nikon S Plan Fluor ELWD 20x/0.45) and two lenses. As a light source, we use a supercontinuum laser (Fianium Whitelase WL-SC-400-2) with a tunable spectral filter (NKT Photonics SuperK VARIA) to obtain a spectral bandwidth of 10 nm. The polarization state of the light was set and after the samples filtered by pairs of a quarter-waveplate and a linear polarizer.

The silicon metasurfaces were fabricated on a glass substrate following the processes of silicon deposition, patterning, lift-off, and etching. First, through plasma-enhanced chemical vapor deposition, we prepared a 600 nm thick amorphous silicon (a-Si) film. Following this, a poly(methyl methacrylate) resist layer was spin-coated and patterned by standard electron beam lithography. After development, a 20 nm thick chromium layer was deposited. After a lift-off process, we used inductively coupled plasma reactive ion etching to transfer the predefined structures by the chromium mask to the silicon. Afterward, the residual chromium mask was removed using a standard wet etching process. Each metasurface hologram has a pattern size of 400x400 µm² and consists of 80x80 pixels yielding a 5 µm pixel size. Each pixel consists of an array of silicon nanofins, which have a grid periodicity of 500 nm. Each cuboid nanostructure has a length of 190 nm and a width of 130 nm.

*Metasurface design*

The phase mask calculations were performed in Python v3.8 with the usage of Tensorflow v2.2 as an automatic differentiation library. For the single nanostructure design, we use rigorous coupled-wave analysis. More details can be found in the Supplementary Information.

## Author contributions

L.H. proposed the conceptional idea, which P.G. and Q.W. refined later. P.G. and Q.W. developed the holographic optimization algorithm. Q.W. and L.H. designed and B.S. fabricated the metasurfaces. P.G., C.S., and T.Z. carried out the optical experiments. Y. W., L.H., and T.Z. supervised the overall projects. All authors analyzed the data, discussed the results, and prepared the manuscript.

## Acknowledgment

The authors acknowledge the funding provided by the National Key R&D Program of China (No. 2017YFB1002900) and the European Research Council (ERC) under the European Union's Horizon 2020 research and innovation program (grant agreement No. 724306). We also thank the NSFC-DFG joint program (DFG No. ZE953/11-1, NSFC No. 61861136010) for additional support. L.H. acknowledges the support from the Beijing Outstanding Young Scientist Program (BJJWZYJH01201910007022), the National Natural Science Foundation of China (No. 61775019) program, and the Fok Ying-Tong Education Foundation of China (No.161009).

## Data availability

All data needed to evaluate the conclusions in the paper are present in the paper and/or the Supplementary Materials. The software codes and the raw data that support the plots within this paper are available from the corresponding author upon reasonable request.

## References

1. A. Shamir, How to Share a Secret. *Communications of the Acm* 22, 612-613 (1979).
2. G. R. Blakley, in *1979 International Workshop on Managing Requirements Knowledge (MARK)*. (IEEE, 1979), pp. 313-318.

3. L. M. Lancaster, The future for security applications of optical holography. *International Conference on Applications of Optical Holography* 2577, 71-76 (1995).
4. L. Kotačka, T. Těthal, V. Kolařík, in *Congress on Optics and Optoelectronics*. (SPIE, 2005), vol. 5954, pp. 59540K.
5. B. Javidi *et al.*, Roadmap on optical security. *Journal of Optics* 18, 083001 (2016).
6. Z. L. Deng, G. X. Li, Metasurface optical holography. *Materials Today Physics* 3, 16-32 (2017).
7. P. Genevet, F. Capasso, Holographic optical metasurfaces: a review of current progress. *Reports on Progress in Physics* 78, 024401 024401 (2015).
8. L. L. Huang *et al.*, Broadband Hybrid Holographic Multiplexing with Geometric Metasurfaces. *Advanced Materials* 27, 6444-6449 (2015).
9. D. D. Wen *et al.*, Helicity multiplexed broadband metasurface holograms. *Nature Communications* 6, 8241 (2015).
10. W. Y. Zhao *et al.*, Full-color hologram using spatial multiplexing of dielectric metasurface. *Optics Letters* 41, 147-150 (2016).
11. K. T. P. Lim, H. L. Liu, Y. J. Liu, J. K. W. Yang, Holographic colour prints for enhanced optical security by combined phase and amplitude control. *Nature Communications* 10, 25 (2019).
12. W. W. Wan, J. Gao, X. D. Yang, Full-Color Plasmonic Metasurface Holograms. *ACS Nano* 10, 10671-10680 (2016).
13. J. X. Li *et al.*, Addressable metasurfaces for dynamic holography and optical information encryption. *Science Advances* 4, eaar6768 (2018).
14. S. M. Kamali *et al.*, Angle-Multiplexed Metasurfaces: Encoding Independent Wavefronts in a Single Metasurface under Different Illumination Angles. *Physical Review X* 7, 041056 (2017).
15. H. Ren *et al.*, Metasurface orbital angular momentum holography. *Nature Communications* 10, 2986 (2019).
16. X. Fang, H. Ren, M. Gu, Orbital angular momentum holography for high-security encryption. *Nature Photonics* 14, 102-108 (2020).
17. P. C. Li, E. T. Yu, Flexible, low-loss, large-area, wide-angle, wavelength-selective plasmonic multilayer metasurface. *Journal of Applied Physics* 114, 133104 (2013).
18. P. C. Li, E. T. Yu, Wide-angle wavelength-selective multilayer optical metasurfaces robust to interlayer misalignment. *Journal of the Optical Society of America B* 30, 27-32 (2013).
19. O. Avayu, E. Almeida, Y. Prior, T. Ellenbogen, Composite functional metasurfaces for multispectral achromatic optics. *Nature Communications* 8, 14992 (2017).
20. Y. Zhou *et al.*, Multilayer Noninteracting Dielectric Metasurfaces for Multiwavelength Metaoptics. *Nano Letters* 18, 7529-7537 (2018).
21. E. Arbabi *et al.*, MEMS-tunable dielectric metasurface lens. *Nature Communications* 9, 812 (2018).
22. A. Arbabi, E. Arbabi, Y. Horie, S. M. Kamali, A. Faraon, Planar metasurface retroreflector. *Nature Photonics* 11, 415-420 (2017).
23. D. Frese, Q. S. Wei, Y. T. Wang, L. L. Huang, T. Zentgraf, Nonreciprocal Asymmetric Polarization Encryption by Layered Plasmonic Metasurfaces. *Nano Letters* 19, 3976-3980 (2019).
24. Y. Zhao, M. A. Belkin, A. Alu, Twisted optical metamaterials for planarized ultrathin broadband circular polarizers. *Nature Communications* 3, 870 (2012).
25. Y. Zhou *et al.*, Multifunctional metaoptics based on bilayer metasurfaces. *Light-Science & Applications* 8, 80 (2019).
26. Y. Q. Hu *et al.*, 3D-Integrated metasurfaces for full-colour holography. *Light-Science & Applications* 8, 86 (2019).
27. Y. Zhou, H. Y. Zheng, I. I. Kravchenko, J. Valentine, Flat optics for image differentiation. *Nature Photonics* 14, 316-323 (2020).
28. H. Kwon, E. Arbabi, S. M. Kamali, M. Faraji-Dana, A. Faraon, Single-shot quantitative phase gradient microscopy using a system of multifunctional metasurfaces. *Nature Photonics* 14, 109-114 (2020).
29. X. Lin *et al.*, All-optical machine learning using diffractive deep neural networks. *Science* 361, 1004-1008 (2018).

30. Y. Gao *et al.*, Multiple-image encryption and hiding with an optical diffractive neural network. *Optics Communications* 463, 125476 (2020).
31. P. Refregier, B. Javidi, Optical-Image Encryption Based on Input Plane and Fourier Plane Random Encoding. *Optics Letters* 20, 767-769 (1995).
32. S. Deng, L. Liu, H. Lang, D. Zhao, X. Liu, Cascaded Fresnel digital hologram and its application to watermarking. *Optica Applicata* XXXVI, 413-420 (2006).
33. G. Qu *et al.*, Reprogrammable meta-hologram for optical encryption. *Nature Communications* 11, 5484 (2020).
34. Y. Z. Li, K. Kreske, J. Rosen, Security and encryption optical systems based on a correlator with significant output images. *Applied Optics* 39, 5295-5301 (2000).
35. G. H. Situ, J. J. Zhang, A cascaded iterative Fourier transform algorithm for optical security applications. *Optik* 114, 473-477 (2003).
36. W. Chen, B. Javidi, X. D. Chen, Advances in optical security systems. *Advances in Optics and Photonics* 6, 120-155 (2014).
37. A. G. Baydin, B. A. Pearlmutter, A. A. Radul, J. M. Siskind, Automatic Differentiation in Machine Learning: a Survey. *Journal of Machine Learning Research* 18, 5595–5637 (2018).
38. Z. Bomzon, V. Kleiner, E. Hasman, Pancharatnam-Berry phase in space-variant polarization-state manipulations with subwavelength gratings. *Optics Letters* 26, 1424-1426 (2001).
39. L. L. Huang *et al.*, Dispersionless Phase Discontinuities for Controlling Light Propagation. *Nano Letters* 12, 5750-5755 (2012).
40. D. J. Lu *et al.*, Experimental optical secret sharing via an iterative phase retrieval algorithm. *Optics and Lasers in Engineering* 126, 105904 (2020).
41. Z. Li, M. Premaratne, W. Zhu, Advanced encryption method realized by secret shared phase encoding scheme using a multi-wavelength metasurface. *Nanophotonics* 9, 3687-3696 (2020).
42. D. P. Kingma, J. Ba, Adam: A Method for Stochastic Optimization. *arXiv*, 1412.6980 [cs.LG] (2014).

# Supplementary Information

## Optical secret sharing with cascaded metasurface holography

Philip Georgi*,1, Qunshuo Wei*,2, Basudeb Sain[1], Christian Schlickriede[1], Yongtian Wang[2], Lingling Huang[2] and Thomas Zentgraf[1]

[1] Department of Physics, Paderborn University, Warburger Straße 100, 33098 Paderborn, Germany
[2] Beijing Engineering Research Center of Mixed Reality and Advanced Display, School of Optics and Photonics, Beijing Institute of Technology, 100081, Beijing, China

### 1. Optical setup for measuring the holographic images

The light of a tunable laser source (Fianium Whitelase WL-SC-400-2) passes through a linear polarizer and a quarter-waveplate to set the polarization state. Next, the light is weakly focused on the cascaded samples and collected by a microscope objective lens. The polarization state is filtered out by the second set of a quarter-waveplate and a linear polarizer. Finally, the Fourier space is imaged by a CMOS camera. The schematic drawing of the setup is shown in Figure S1.

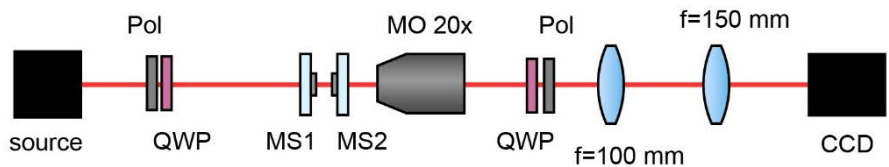

Figure S1: Schematic of the experimental setup for measuring the holographic images. Pol - linear polarizer, QWP – quarter waveplate, MS – metasurface hologram, MO – microscope objective.

In our experimental setting, the usage of polarization optics allows us to partially filter out unwanted artifacts in all measured Fourier images (see Figure 2 and Figure 4). For the single-layer images, the transmitted co-polarized light will cause a zeroth order, which will be filtered out with polarization optics.

For the cascaded hologram, things are quite different. Since the circularly polarized light passes through a metasurface twice, its polarization should ideally be converted back and forth. Thus, we need to measure in co-polarization in our experiment. In the cascaded measurements, there are two types of errors that can occur: First, some light components might only switch their circular polarization state once, which would correspond to the single images. Fortunately, we measure in the co-polarization and filter out this case. Second, some light parts might pass through both metasurface without changing their circular polarization state, which leads to a zeroth-order spot. Unfortunately, we are unable to filter out this case since we need to measure in co-polarization.

## 2. Alignment sensitivity for the cascaded images

In all used measurement settings, the image formation for the cascaded images is highly alignment dependent. The following measurement results quantify this alignment sensitivity with regard to translational movements. In the first experimental case, we observe that the cascaded image becomes unrecognizable when the one metasurface (in Set A) is moved 10 µm (2 pixels) in the x-direction (see Figure S2). This high alignment sensitivity for in-plane movement can be explained by the requirement of pixel matching. If both phase-masks would be placed in the same plane a one-pixel movement would completely erase the cascaded image. However, since there is a propagation distance between both metasurfaces, we have some interaction with neighboring pixels between the two metasurfaces, which softens the alignment requirement.

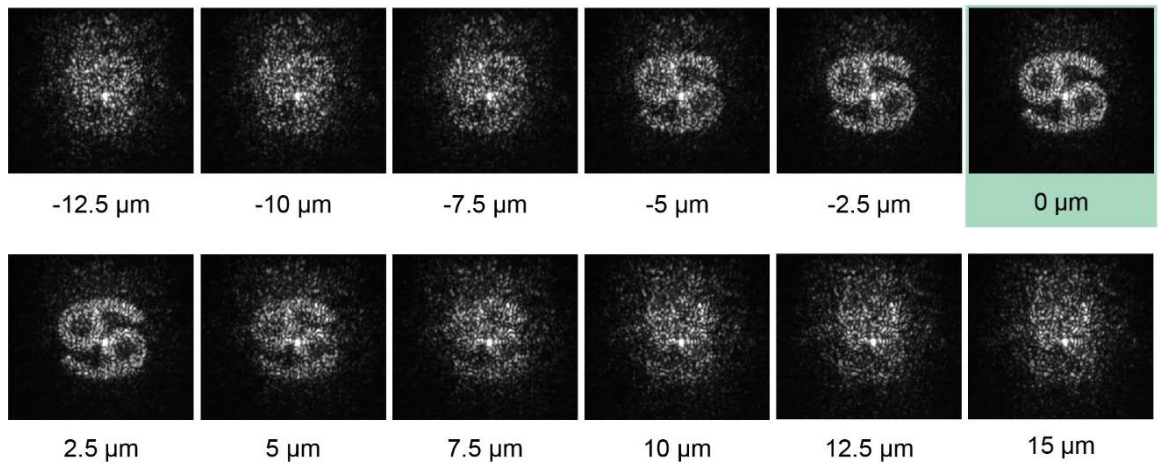

Figure S2: Cascaded image formation for a translational in-plane movement (x-shift) of one metasurface (set A) in the first experimental case. The green marker highlights the configuration with the sharpest image formation.

The high alignment sensitivity with regard to the in-plane movement was something that we anticipated in advance and that was confirmed in early simulations before the actual metasurface fabrication. Therefore, we intentionally choose to fabricate a pixel size larger than the intrinsic metasurface resolution to increase to pixel overlap area in the case of misalignment. As a side effect, the higher pixel size should also lead to a decrease in diffraction strength for the light propagation from one metasurface to the other. Thus, the cascaded setup becomes overall less alignment sensitive for a higher pixel size at the cost of lower image qualities due to less available pixels as well as the appearance of higher diffraction orders. It should be noted, that spatial light modulators (SLMs) have similar pixel sizes as the used 5 µm and thus present a real alternative. However, in practice, it might be very difficult to achieve small distances between two commercial SLMs due to mechanical blockage.

In our experiments, we found that the cascaded image formation is a lot less sensitive towards translational movements out of the plane (z-direction) than to in-plane movements (x-direction). Thus, we roughly have to move the first metasurface 500 µm further away from the second metasurface to observe a deterioration of the cascaded image (see Figure S3). In our experimental setting, a minimum distance of 60 µm can be achieved. In principle, the two metasurfaces can be brought arbitrarily close together. It is worth noting that the minimum distance achievable is zero, which can be realized with the help of 4f system. However, using 4f system will introduce optical aberrations and therefore limits the achievable resolution, while direct cascading two metasurfaces is more convenient and intuitive. Without using a 4f system, the distance between the two metasurfaces is mostly limited by the experimental setup. For very small distances, unwanted optical near-field interactions like Fabry Pérot modes can appear.

In our experiments, we did not observe any obvious Fabry Pérot effects when changing the metasurface distance. This circumstance can be explained if we calculate the free spectral range in terms of the wavelength of a Fabry Pérot Interferometer with a cavity length of 100 µm:

$$\delta\lambda = \frac{\lambda^2}{2L} = \frac{(800\text{ nm})^2}{2 \cdot 100\text{ µm}} = 3{,}2\text{ nm}$$

Since this value is smaller than the actual bandwidth of incident light which was 10 nm, any Fabry Pérot effect should average out over the used spectral range in the final image. Since the metasurface was designed for the operation in transmission, we expect a rather small influence of multireflections in the final image.

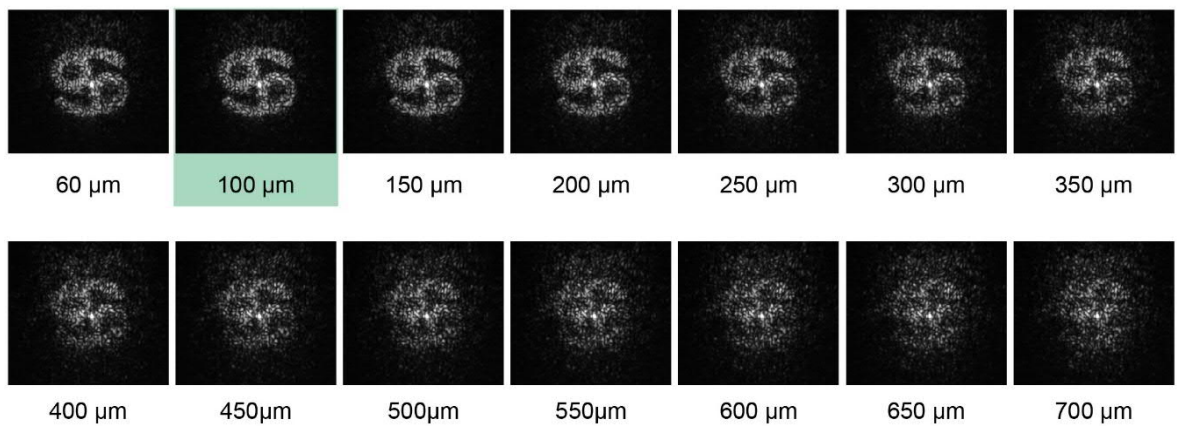

Figure S3: Cascaded image formation for different metasurface distances in the first experimental case. During this experiment, the first metasurface (set A) is moved away from the second metasurface (z-shift). The cascaded image is sharpest for a metasurface distance of 100 µm (green marker).

In terms of alignment sensitivity all three demonstrated cases behave in the same way. Figure S4 shows the transition from the cascaded image '3' to the cascaded image '4' by translational in-plane movement (x-shift). The measured translational shift between both sharp images is consistent with the designed shift of 25 µm (5 pixels). A notable observation is, that there is no organic transition like in a humanly produced animation. Instead, we see a fading transition with a noisy pattern in between. Compared to the first experimental case (Figure S2), the image deteriorates slightly slower. However, it is uncertain, whether this difference stems from the algorithmic design or if it is caused by an alignment uncertainty.

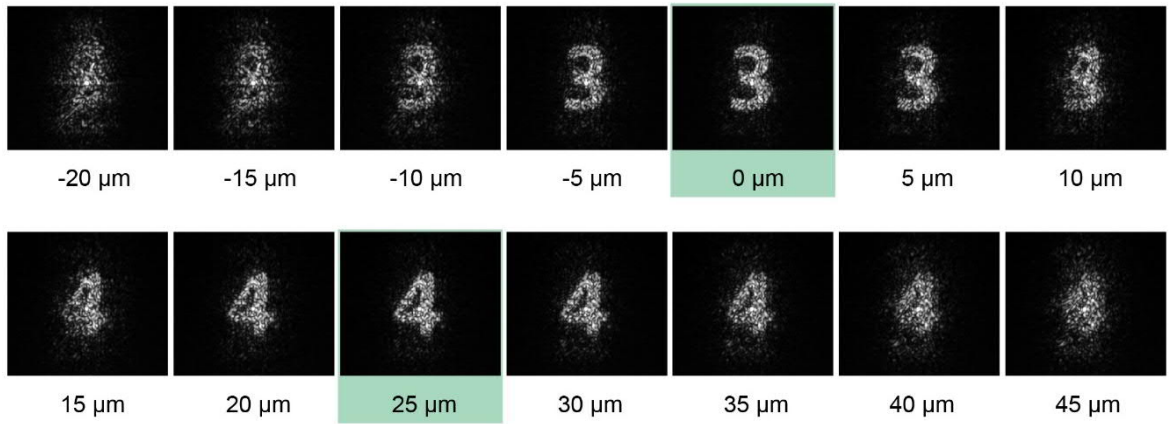

Figure S4: Cascaded image formation for the translational shift from the image '3' to '4'. We observe that the translational shift between the measured sharp images does correspond to the designed 5 pixels shift (25 μm).

## 3. Optical secret splitting with polarization multiplexed metasurfaces

Our demonstrated cascaded encryption scheme is further expandable. In the following, we present an additional framework, in which we combine established metasurface polarization multiplexing concepts with the encryption scheme shown in Figure 2 of the main manuscript. The usage of polarization multiplexing techniques offers an increased information capacity per hologram at the cost of being more reliant on a high-quality fabrication. In the here presented framework Set A consists of a single polarization-independent metasurface hologram, while Set B consist of two polarization-multiplexed metasurface holograms, that utilize different multiplexing methods (see Figure S5).

The basic idea is that each metasurface in Set B is essentially carrying two different phase masks, which can be selected according to the choice of polarization in the setup. Thus, we have effectively one hologram in Set A and four holograms in Set B, which allows for the encoding of five single-layer images four cascaded images. As encoded images, we choose different single letters. These letters are positioned in such a way the overlap of a cascaded image with its two according single-layer images create a 3-letter word (Figure S5).

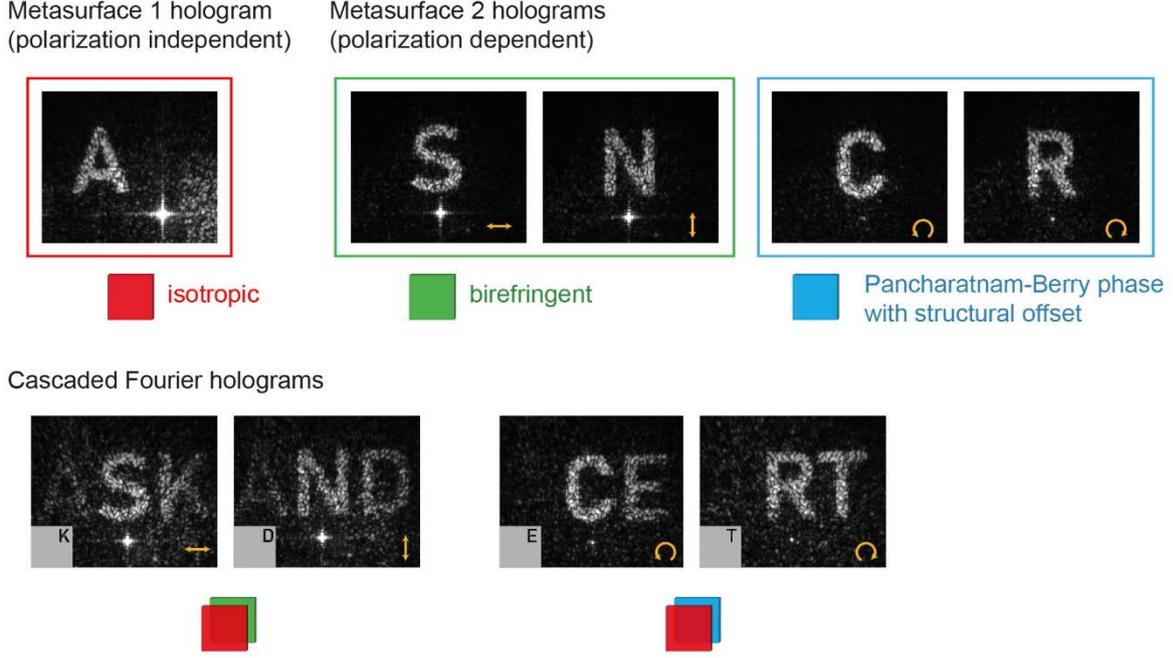

Figure S5: Polarization multiplexed encryption. In our polarization multiplexing scheme, we cascade a polarization-independent metasurface (red) with a polarization multiplexed metasurface (green/blue). An overview of all used nanostructure designs is presented in Table S1. Here, all measured holographic images are shown for the corresponding input polarization (yellow arrows). Note, that all measured cascaded holograms contain (unwanted) relics of their associated single-layer images due to optical discrepancies between the design and fabrication of the metasurfaces, which are difficult to avoid. In the ideal case, the cascaded images would only include the last visible letter as shown in the grey target images.

Due to their different functionalities, all metasurface holograms use different nanostructure design concepts. The polarization independent metasurface (red) consists of differently-sized square-shaped structures. Due to their symmetry, these nanostructures inherently only transmit light in the co-polarization for any input polarization, which carries a unique phase based on the chosen structure size.

The birefringent metasurface (green) consists of rectangular-shaped nanofins. Since these structures look different across x- and y-direction, horizontally and vertically polarized light accumulate different phases on transmission. Thus, we can encode two different phase masks within the same metasurface hologram.

The third metasurface design (blue) is based on the Pancharatnam-Berry phase $\phi_{PB}$. Thus, each rectangular-shaped nanofin is designed as a local half-wave plate, which introduces a geometric phase according to the fin's orientation into the transmitted light for a circular input polarization. However, by using different nanofin shapes, we can encode an additional shape-dependent offset phase $\phi_{structure}$, which allows for the encoding of two independent phase masks within the same metasurface:

$$\phi_{RP/LP} = \phi_{structure} \pm \phi_{PB}$$

| | | | |
|---|---|---|---|
| SEM image | 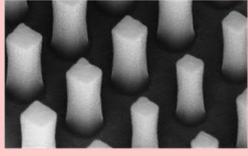 | 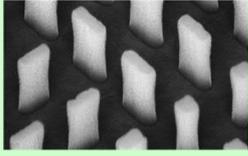 | 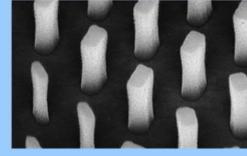 |
| Shape | Squared | Rectangular | Rectangular |
| Rotation | No | No | Yes |
| Input polarization | Any | Linear | Circular |
| Transmitted polarization | Co-polarized | Co-polarized | Cross-polarized |
| Polarization dependence of the transmitted phase | No | Yes | Yes |

Table S1: Overview of the different nanostructure designs for the polarization multiplexing framework. The colors are chosen according to Figure S5.

In our measurement, all single metasurface holograms are clearly visible (Figure S5 top row). However, for the metasurfaces measured in co-polarization (red/green), we observe a considerable zeroth-order spot. This zeroth-order spot is usually the result of systemic deviations between the fabricated metasurface and its ideal phase mask. For an average Fourier hologram, we would usually expect, that all phase values are appearing roughly equally frequent. Thus, the average transmission coefficient, which corresponds to the zeroth-order intensity is near zero. However, if the fabricated metasurface introduces some systemic errors (with regard to refraction indices or structural sizes), the metasurface might not achieve all targeted phase values equally efficient. Thus, the average transmission coefficient for the metasurface shifts away from zero and we observe a high zeroth-order intensity. In the cross-polarization case (blue metasurface) this zeroth-order spot also exists, but it is filtered out by polarization filters.

In the cascaded images we additionally observe that the measured images are also containing the letters of their respective single images. The main cause for this is the abovementioned systematic errors between fabrication and design. In a simple picture, we could consider each metasurface as a superposition of the target phase mask and a constant phase mask. Light passing through both metasurface becomes a superposition of all phase mask combinations. Thus, the cascaded image can contain all four possible combinations: the cascaded letter, both single letters & the zeroth-order spot. For the first cascaded case (red and green metasurface), it is notable that the letter 'A' is considerably weaker than the letter 'S'. This observation indicates, that the birefringent metasurface (green) is overall more efficient than the polarization-independent metasurface (red). This is consistent with the circumstance that the zeroth-order spot is greater in the single-layer image of the first metasurface (red). In the second cascaded case, the letter 'A' does not appear at all as it is filtered out by polarization filters.

## 4. Nanostructure design

For the design and optimization of the fabricated nanostructures, we use the simulation method of rigorous coupled-wave analysis (RCWA). As a general blueprint, we work for all produced metasurfaces with cuboid-shaped silicon nanofins, which are fabricated on top of an ITO glass substrate in a squared shaped lattice. Since RCWA only allows for periodic structures, we simulate all single nanostructures only in this periodic environment.

For our simulation, we choose a fixed nanofin height of 600 nm with a period of 500 nm in both x- and y-directions and a wavelength of 800 nm. We swept the nanofin length and width in a range of 70 nm to 300 nm in 5 nm steps, which provides us with a 2D-map of available nanostructures for all used metasurface designs.

In our experiments, we worked with different metasurface design concepts. The most important concept is the Pancharatnam-Berry phase, which was exclusively used for the work shown in our main manuscript. For the Pancharatnam-Berry phase, we designed a structure, which behaves like a local half waveplate and maximizes the circular cross-polarization transmittance. The simulated amplitude of the cross- and co-circularly polarized transmission coefficients of the nanofins are shown in Figure S6a and Figure S6b. From the simulation results, one can directly select the structures with high cross circularly polarized transmittance and low co circularly polarized transmittance at the working wavelength of 800 nm. Considering the balance of fabrication accuracy and the broadband property, we chose a length of 190 nm and a width of 130 nm for the PBP-only metasurface (marked as a white point). For the polarization multiplexing case, we require an additional structural offset phase. Thus, we choose multiple geometries that cover the whole $2\pi$- range for the offset phase (see Figure S6c).

The second type of metasurface, which we used is the birefringent metasurface. Figure S7 shows the simulated amplitude and phase of the transmission coefficients for x-polarization and y-polarization. From these plots, one can see that the phase modulation varies quickly and can cover the phase range from 0 to $2\pi$ smoothly while the amplitude remains high. Therefore, we can easily select the target structures with the desired phase modulation for x- and y-polarization channels, simultaneously. From the structure size versus frequency plot (Figure S7c), we can see that the most used nanofin structure sizes have a length and width around 150nm. In this area, the $t_{xx}$- and the $t_{yy}$-phases are highly size-dependent.

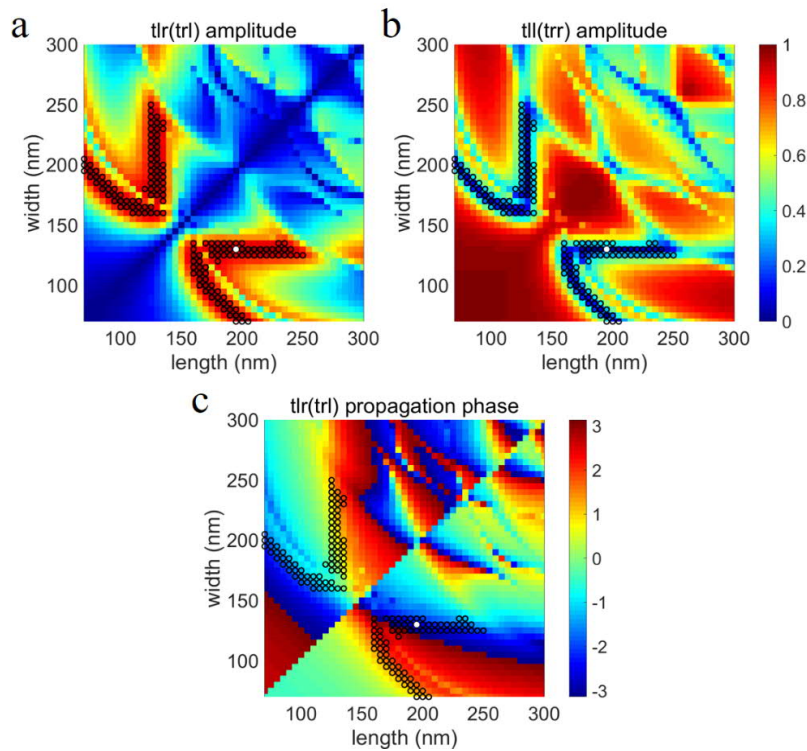

Figure S6: Pancharatnam-Berry phase based nanostructure design. a and b, Amplitude map of the circular transmission coefficient in cross- and co-polarization for different geometry sizes. c, The additional shape-dependent offset phase of the circular cross-transmission coefficient for different geometry sizes. In all plots, the white point refers to the geometry in the main script and the black circles refer to the structure sizes used in the polarization multiplexing case.

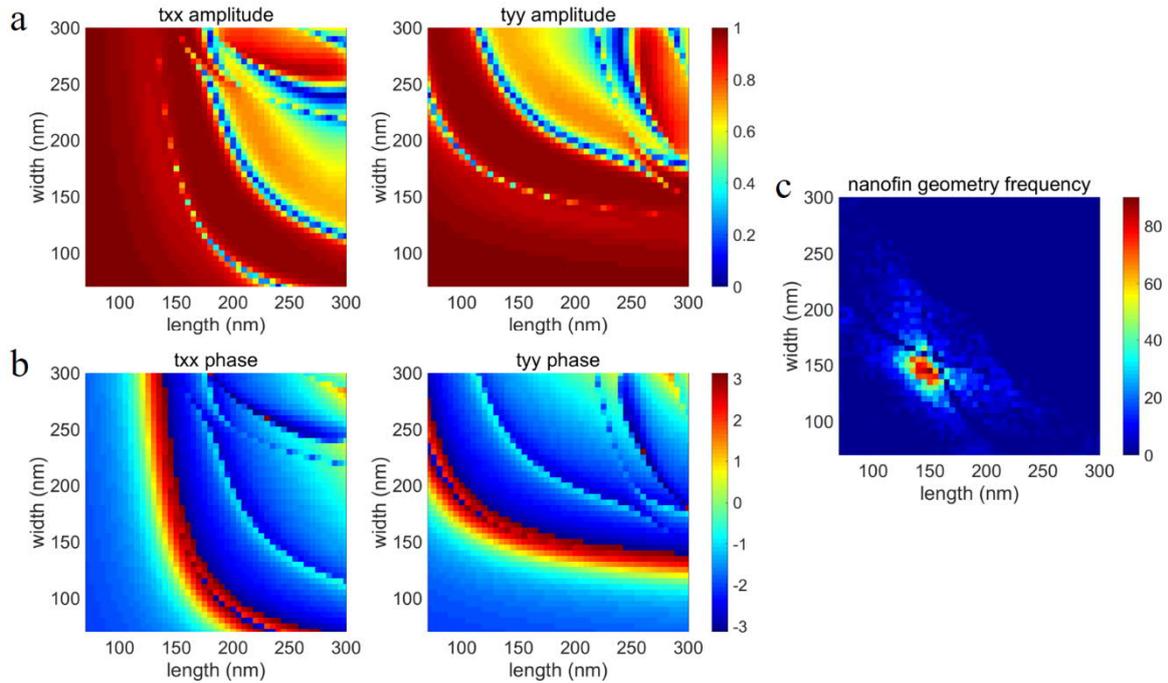

Figure S7: Birefringent metasurface design. a and b, amplitude and phase maps for all linear transmission coefficients. c, statistical frequency of the different nanofin geometries that are placed across the birefringent metasurface hologram.

The last used metasurface design is the isotropic design, which only allows for squared shaped nanostructures. In Figure S8 the phase and amplitude of the transmission coefficient are plotted over a side length range between 70 nm and 190 nm. By selecting those isotropic structures within this range, we can cover the full $2\pi$ phase range while keeping the amplitude almost uniformly with an average above 95%. Only at the resonance dip for a side length of 155 nm, these isotropic nanostructures become considerably less efficient.

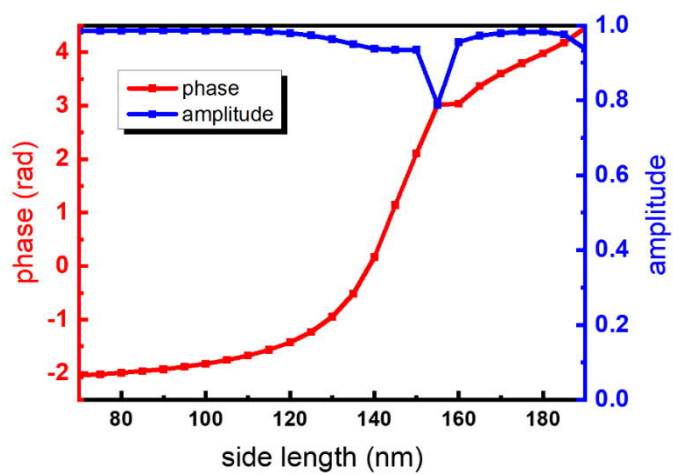

Figure S8: Isotropic metasurface design. Phase and amplitude are shown over the used side length range, which covers a full $2\pi$ shift.

## 5. Spectral efficiency measurement

We used the wavelength as an additional degree of freedom to compensate for systematic deviations between design and fabrication. Thus, we measured the transmission of the metasurface for circular co- and cross-polarization for multiple wavelengths. The thereby obtained relative conversion efficiency is slightly shifted to shorter wavelengths in comparison to the calculated spectrum (see Figure S9). Thus, we set the measurement wavelength at the measured efficiency maximum at 740 nm. For the polarization multiplexed case such an approach is not available. Instead, we choose the wavelength which leads to the subjectively best image qualities, which was around 700 nm.

The shifted transmission in the experiment to shorter wavelengths compared to the design can be the result of small deviations of the refractive index of the silicon from the design values as well as from the deviation of the structure dimensions (in particular the height). Numerical simulations show that a deviation of 5% from the design values of dimensions can result in a shift of 30 nm to shorter wavelengths.

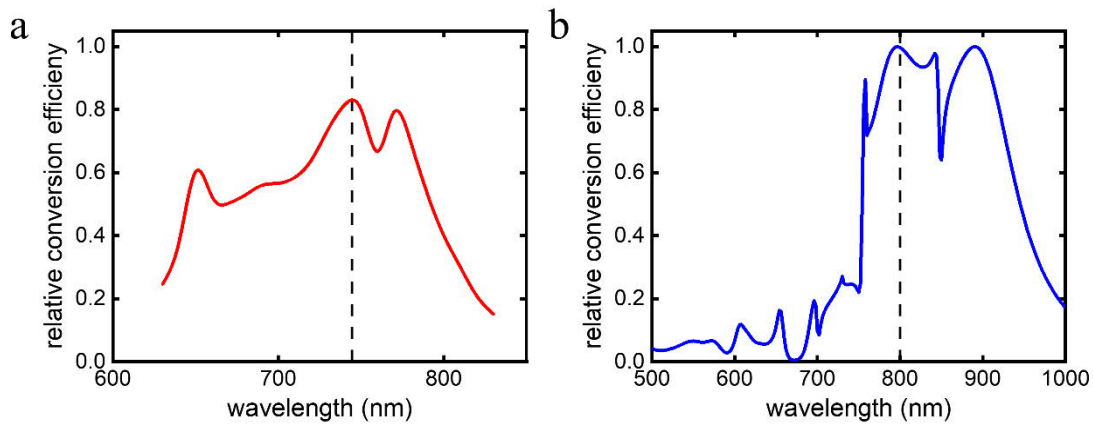

Figure S9: Efficiency of the Pancharatnam-Berry metasurface. a, measured spectral relative conversion efficiency ($T_{lr}/(T_{lr}+T_{rr})$). b, simulated spectral relative conversion efficiency. The simulation results are calculated by the method of rigorous coupled-wave analysis (RCWA). The dashed lines in a and b show the design and the measurement wavelength (740 nm and 800 nm). Notably, the fabricated metasurface is most efficient at a shorter wavelength in comparison to its simulated design.

## 6. The gradient optimization scheme

In its core, the gradient optimization scheme consists of two different parts: the explicit formulation of the forward calculation as well as the usage of a gradient optimizer. Here, we will start with the description of the used optimizer. In the research area of machine learning, the most popular gradient optimizers are stochastic gradient descent (SGD) and Adam(*1*). While the SGD simply moves along the calculated gradient, Adam additionally keeps track of previous gradients and prioritizes changing the variables with insensitive gradient components (low noise) to achieve faster convergence. However, for both gradient optimizers the choice of the hyperparameter 'learning rate', which scales the step size per iteration, heavily determines the convergence behavior. If the learning rate is set too high, the optimization scheme does not converge at all and if the learning rate is set too low, convergence becomes very slow. To circumvent this issue, we introduce a novel gradient optimizer based on Adam which adjusts the learning rate during the optimization. Thereby, the optimization process becomes resilient towards its hyperparameters and we can guarantee a convergence towards a local optimum.

The underlying idea for our new optimizer is quite simple: For a given phase mask set, the Adam optimizer suggests a new point in each iteration. However, instead of just choosing that point and repeating the process, we take the same step once more and make a second evaluation. (The second point corresponds to a single step with a doubled learning rate.) We then compare the initial point with the two new candidates and choose the point with the lowest associated error. Additionally, we modify the learning rate according to the comparison result (see Table S2). Thus, we slow down, when we overshoot a local minimum in the subspace and accelerate otherwise.

| evaluated point with the lowest error | sketch | change of the learning rate |
|---|---|---|
| double step | 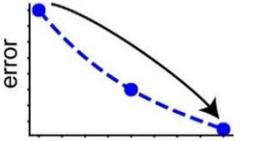 | increase learning rate (factor 1.1) |
| one step | 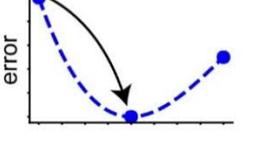 | decrease learning rate (factor 0.7) |
| original point | 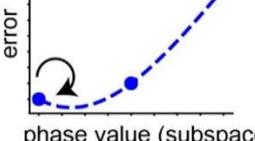 | decrease learning rate (factor 0.1) |

Table S2: Adaptive learning rate method. By evaluating 2 points along the gradient or momentum direction and comparing them with the starting point, the learning rate can be adaptively changed during the optimization process to speed up convergence.

The formulation of the forward pass is changing for the different experimental cases. In the main manuscript, the forward pass for the first experimental case is explained. For the polarization multiplexing case (Figure S5), we use the same code for the optimization process. Thereby, we just treat the polarization multiplexed metasurfaces as two independent phase masks. For the translational multiplexed case (Figure 4), we need to slightly modify the forward pass simulation. As demonstrated in Figure S10, this involves the previously described periodic expansion of the first phase mask, as well as a cropping operation according to the relative translational position.

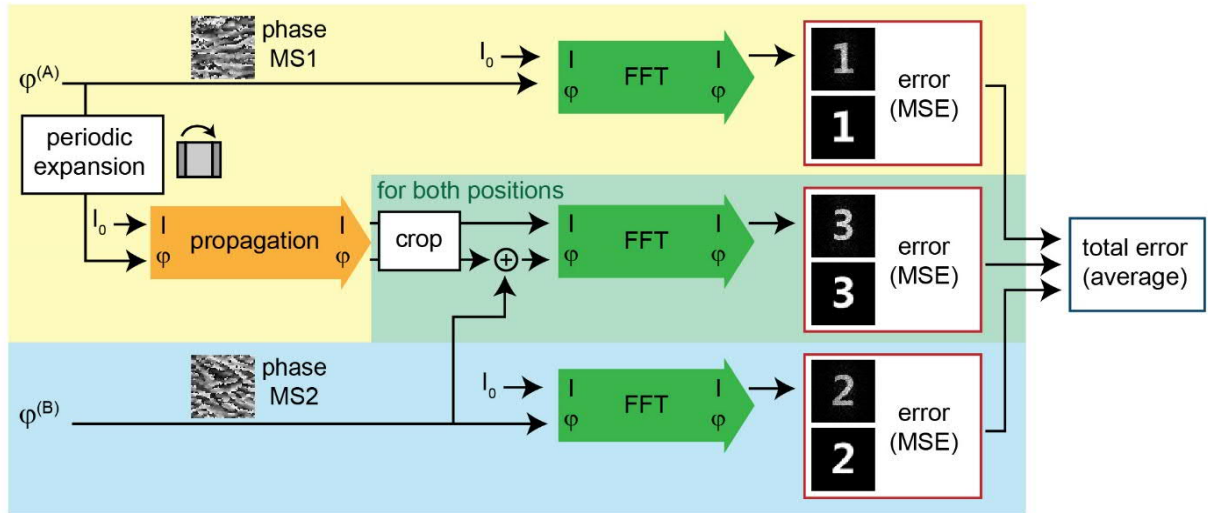

Figure S10: Forward pass for the translational case. In comparison to the first experimental case, two things change: For the recreation of the cascaded images we first periodically expand the quadratic phase mask into a rectangle, which corresponds to the fabricated metasurface. Afterward, we crop the propagated light field according to the relative translational position for both cases. The cascaded image creation is then analog to our initial case.

After explaining the functionality of our algorithm, we now want to provide some benchmarking. For the first experimental case, we roughly need 0.4 s for each iteration on a moderately equipped PC (Intel i5-7500) and converge after around 200 iterations (see Figure S11). This comparably fast calculation time for a hologram generation is connected to the overall low pixel count.

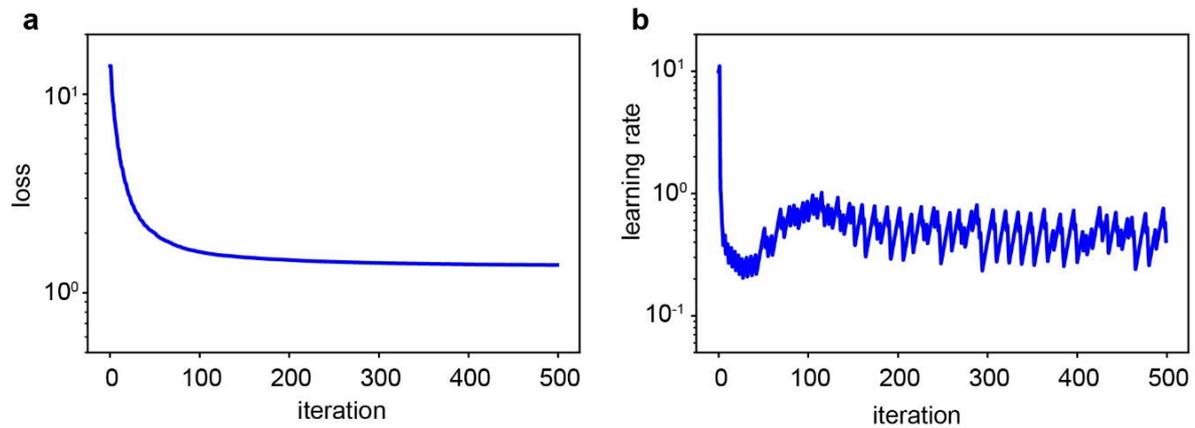

Figure S11: Convergence plots for the gradient optimization scheme in the first experimental case. a, The loss, which is the calculated error in each iteration (Figure 3), benchmarks how well the current phase mask sets create all images. We observe that after around 200 iterations convergence has been achieved. b, The learning rate controls the step size during the optimization process. Our adaptive mechanism quickly turns down our initially chosen learning rate value to a more and then adjusts it dynamically according to the optimization state. However, the systematic changes of the learning rate during the optimization are difficult to understand, since many factors play a role like a gradient strength, the gradient stability, local nonlinearity, and the distance to the local minima.

To quantitatively estimate the quality of the reconstructed images, the correlation coefficients between reconstructed images and target images are analyzed as the criterion. The correlation coefficients can be express as

$$cc(T,R) = \frac{COV(T,R)}{\sqrt{D(T)}\sqrt{D(R)}}$$

Where *T* and *R* represent the amplitudes of the target object and reconstructed image, *COV(T, R)* is the covariance of *T* and *R*, *D(T)*, and *D(R)* represent the variances of *T* and *R*, respectively.

The simulated reconstructed astrological symbols in the first experimental case and their correlation coefficients are shown in Figure S12. All the correlation coefficients are larger than 0.95, which is a very high threshold value (maximum 1). Thus, the gradient optimization scheme delivers high-quality holograms despite the high dimensional optimization task. Although it is not guaranteed that the global minimum during the optimization is reached, this is not an issue in the case of our holography based optical encryption scheme (converging only to a local minimum). As long as the total error α is small enough, no matter if it is a local or a global minimum, the individual single-layer holographic image of each metasurface and the cascaded holographic images for the various combinations of the metasurfaces of both sets can be reconstructed with high quality and low cross-talk.

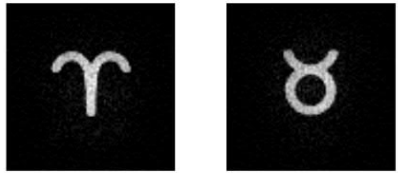
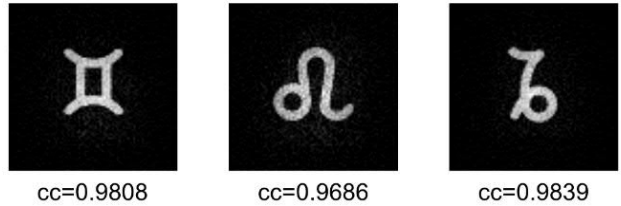
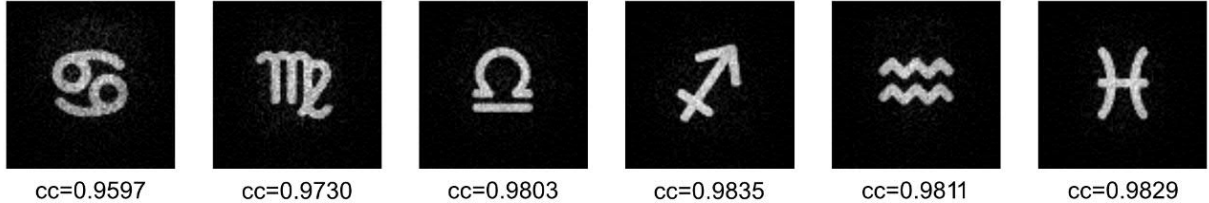

Figure S12: Simulated reconstructed astrological symbols and their correlation coefficients (cc) for the first experimental case.

We also present the optimization results for the case of the translational multiplexing cascaded holograms in Figure S13. In terms of achieved correlation coefficients, the results are similar to the first experimental case, which could have been expected based on a similar phase mask to image ratio.

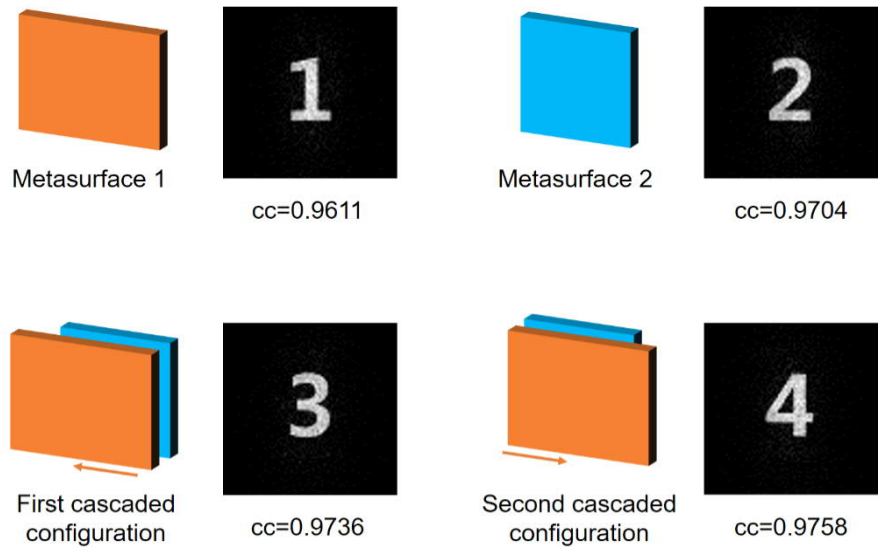

Figure S13: Simulated reconstructed images with their corresponding correlation coefficients for the case of translational multiplexing.

Especially interesting are the simulation results for the polarization multiplexed case in Figure S14. In contrast to the experimental results, there is no crosstalk in the simulated cascaded images which is consistent with the other cases. However, something that we had not predicted is the distribution of the correlation coefficients. While the letter 'A' of the single hologram in Set A was only reconstructed with a correlation coefficient of 0.82, all the other reconstructed images have a correlation coefficient higher than 0.95. The underlying reason might be that the single metasurface hologram in Set A is used for the reconstruction of more images than a phase mask in Set B. Thus, for the chosen metric, it seems to be beneficial to sacrifice some image quality in single-layer images of the smaller set to achieve improvements in the other reconstructed images. However, the differences in image quality could be avoided by changing the relative weighting of errors in the loss calculation in Figure 3.

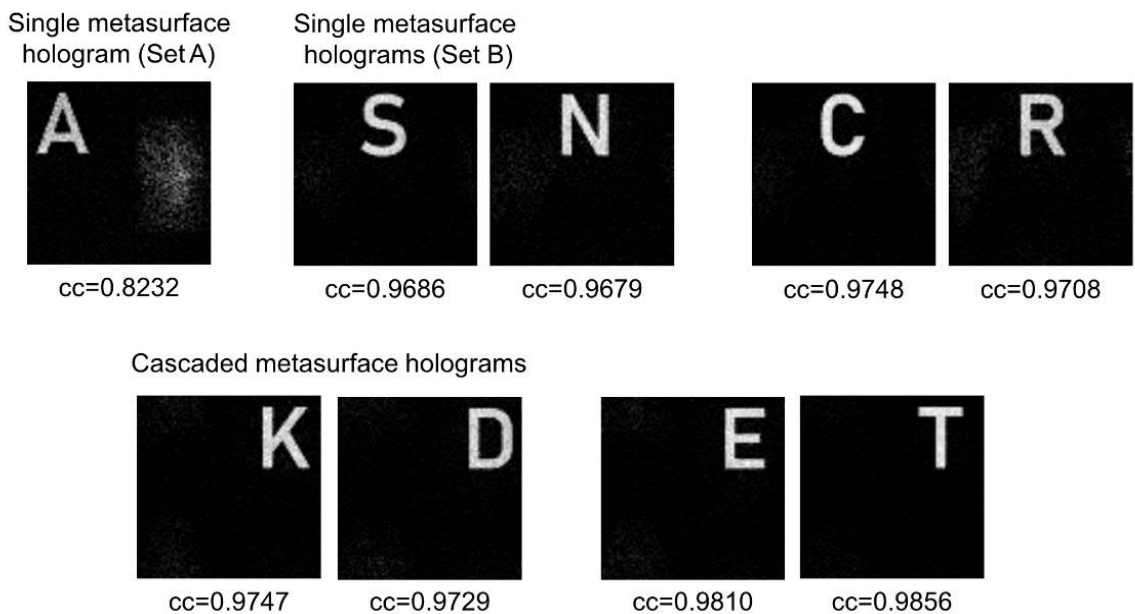

Figure S14: Simulated reconstructed images and corresponding correlation coefficients for the polarization multiplexed cascaded holograms.

## 7. Discussion on cryptographic security strength

In the simplest case, with only two phase patterns, a distance of zero, with one cascaded image and no single layer image, it is actually trivial to prove that our secret sharing scheme is fully secure. In that case, every cascaded phase pattern can still be created for any known phase mask. Thus, the ownership of one phase mask provides absolutely no information about the cascaded image. In other words, the reveal of one phase mask does not change the Shannon-entropy for the second phase mask.

Unfortunately, this security guarantee cannot be extended towards our experimental secret sharing scheme. The issue is, that by simultaneously optimizing for other images (that appear for other combinations and in the single-layer configuration (see Fig. 2)), we effectively obtain new boundary conditions for our phase masks, which might create exploitable correlations. However, this does not mean that our method is insecure. We are currently unaware of any method that breaks our encryption scheme. It is also an intrinsic advantage that for any Fourier hologram the phase mask is not unique for a given holographic image. Thus, even if we had one phase mask and the reconstructed images, we could still not retrieve the second phase mask.